\documentclass[%
 aip,
 amsmath,amssymb,
reprint, onecolumn 
]{revtex4-1}

\usepackage{graphicx}
\usepackage{dcolumn}
\usepackage{bm}

\usepackage[utf8]{inputenc}
\usepackage[T1]{fontenc}
\usepackage{mathptmx}
\usepackage{etoolbox}
\usepackage{xcolor}

\inputencoding{latin1}

\inputencoding{utf8}

\makeatletter
\def\@email#1#2{%
 \endgroup
 \patchcmd{\titleblock@produce}
  {\frontmatter@RRAPformat}
  {\frontmatter@RRAPformat{\produce@RRAP{*#1\href{mailto:#2}{#2}}}\frontmatter@RRAPformat}
  {}{}
}%
\makeatother
\begin{document}
\preprint{AIP/123-QED}

\title{On a Solution to the Dirac Equation with a Triangular Potential Well}

\author{Renebeth B. Payod}
 \email{rbpayod@up.edu.ph}
 \affiliation{Institute of Mathematical Sciences and Physics, University of the Philippines, Los Ba\~nos, Laguna, $4031$ Philippines}

\author{Vasil~A.~Saroka}
 \email{vasil.saroka@roma2.infn.it}
\homepage{https://sites.google.com/view/vasilsaroka/home}
 \homepage{https://www.tbpack.co.uk/}
 \affiliation{Department of Physics, University of Rome Tor Vergata and INFN, Via della Ricerca Scientifica $1$, Roma $00133$, Italy}
 \affiliation{Institute for Nuclear Problems, Belarusian State University, Bobruiskaya 11, 220030 Minsk, Belarus}
 \affiliation{TBpack Ltd., 27 Old Gloucester Street, London, WC1N 3AX, United Kingdom}

\date{\today}

\begin{abstract}
Chiral anomalies resulting from the breaking of classical symmetries at the quantum level are fundamental to quantum field theory and gaining ever-growing importance in the description of topological materials in condensed matter physics. Here we present analytical solutions of the Dirac equation for massless $3+1$ fermions confined to an infinite stripe and placed into a background gauge field forming a triangular potential well across the width of the stripe. Such an effective $1+1$ system hosts zero-energy modes resulting in the gauge field-dependent chiral anomaly structure. This problem has a direct relation to a half-bearded graphene nanoribbon placed into an in-plane external electric field and offers it an exact solution in terms of new special functions that are similar but not reducible to Airy functions.
\end{abstract}

\maketitle

\section{\label{sec:Introduction}INTRODUCTION}
The behavior of electrons in triangular potential wells is described by fundamental solutions known as Airy function~\cite{landau2013quantum,griffiths2018introduction}. Airy functions play an important role in accurately modeling charge carrier properties within two-dimensional electron gases semiconductor devices~\cite{yoshida1986classical,khondker1987analytical}. Understanding electronic transport in semiconductor heterostructures is crucial for designing advanced optoelectronic devices.

At the turn of the century, an increasing number of exact solutions to the Schr{\"o}dinger equation has been reported by using both conventional approach~\cite{zhang1993new,parfitt2002two} and factorization of supersymmetric quantum mechanics (SUSY) ~\cite{lahiri1990supersymmetry,sokolov2008factorization}. Likewise, other solvable systems with exact solutions to the Schr{\"o}dinger equation have been shown in the bound states of a hyperbolic double-well potential~\cite{downing2013solution} and hyperbolic asymmetric double well potential~\cite{hartmann2014bound}. Generally, the analytic solutions to the Schr\"{o}dinger equation were obtained through reduction to a hypergeometric equation~\cite{manning1935exact} or by Heun's differential equation~\cite{downing2013solution,hartmann2014bound,shahnazaryan2014new,ishkhanyan2017}.

Exact solutions of the relativistic Dirac equation supplemented with various background potentials is being a subject of great interest too~\cite{abe1987simple,hofer1989dirac,chargui2010exact,tezuka2013analytical,jaronski2021linear}. Approximate and exact analytical solutions have been reported for a number of cases, including linear scalar and vector potentials~\cite{tezuka2013analytical}. Some of these analytical solutions reduce to Bessel and Airy functions~\cite{jaronski2021linear}. 

\begin{figure}
\includegraphics{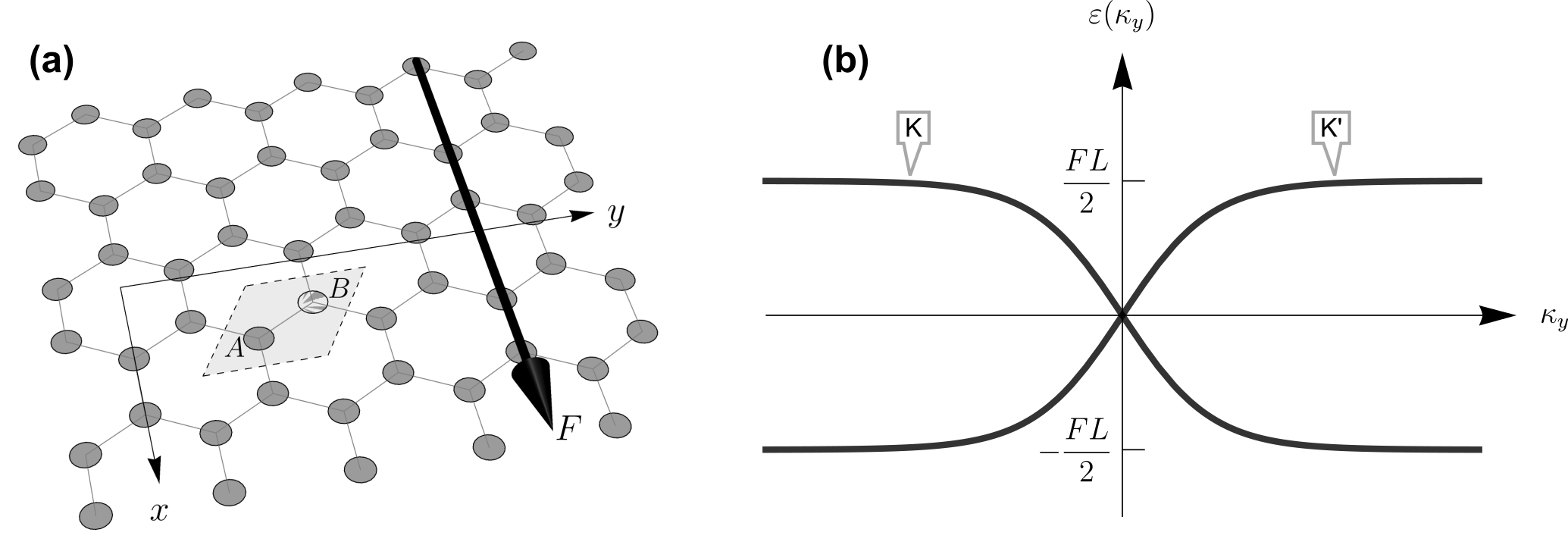}%
\caption{\label{fig:HoneycombLatticeChiralAnomaly} (A) A honeycomb lattice confined to a half-bearded ribbon by cuts normally to $x$-axis at $\pm L/2$ -- hbGNR. Large black arrow shows the direction of the applied in-plane electric field $F$. Gray dashed rectangle is the honeycomb lattice unit cell with two basis atoms (sublattices) denoted as A and B. (B) Gauge field dependent chiral anomaly structure of the energy bands $n=0$.}%
\end{figure}

At the low energy limit in condensed matter systems, the Schr{\"o}dinger equation manifests as a Dirac equation~\cite{jackiw1983continuum,divincenzo1984self}. In two dimensions this happens around degeneracy points in the energy bands, topologically acting as the Berry curvature sources in $k$-space and resembling Dirac monopoles~\cite{renan2000treating,geim2007rise,bergman2009near}. The Berry curvature monopoles in graphene are the ${\bf K}$ and ${\bf K}^{\prime}$ Dirac points. The effective Dirac equation of graphene is capable to describe the electronic properties of graphene nanoribbons (GNR) with different edge structures, such as the prominent zigzag and armchair edge geometries~\cite{brey2006electronic,akhmerov2008boundary}. Despite previous studies done on the edge states and flat bands of graphene nanoribbons~\cite{wakabayashi2010edge,Jaskolski2011,li2012electronic,Saroka2015,Saroka2015a,Saroka2016a,Wakabayashi2001,Kusakabe2003,Kohmoto2007}, a Dirac equation based description is lacking for a GNR known as the half-bearded graphene nanoribbon (hbGNR). Here, we present an analytical treatment of the Dirac equation motivated by hbGNRs and their prominent zero-energy modes~\cite{Wakabayashi2001,Kusakabe2003,Kohmoto2007,saroka2023tunable}. We consider a more general case accounting also the influence of an external in-plane electric field as depicted in Fig.~\ref{fig:HoneycombLatticeChiralAnomaly}(a).

We report here a new set of special functions $R$ and $V$ in the solution of the Dirac equation, which are somewhat analogous but not reducible to the Airy functions. Using a series of substitutions and the power series method, we demonstrate that an exact solution to the Dirac equation relevant to hbGNRs subjected in an external in-plane electric field can be achieved. By combining these analytical solutions with the appropriate boundary conditions, a dispersion equation can be constructed. The dispersion reveals a field-dependent $1$+$1$ chiral anomaly.

\section{\label{sec:Analytical solutions}Analytical solutions for confined massless Dirac fermions coupled to a gauge field.}

We aim to formulate the problem in its most general Lagrangian form to potentially extend its application beyond its original scope. We start from the action of a massless fermion in $3+1$ space-time, $x = \left(ct, x,y,z\right)$, coupled to a gauge field:
\begin{equation}
    S = \int d^4 x\, \mathcal{L} = \int d^4 x \, \bar{\Psi} i \gamma^\mu\ D_{\mu} \Psi , \, 
    \label{eq:action}
\end{equation}
where covariant derivative $D_{\mu} = \partial_{\mu} - i q A_{\mu}$, with $A_{\mu} = \left(\varphi, A_x, A_y, A_z\right)$ being a gauge field represented by $4$-vector of the electromagnetic field, and the summation over repeated index $\mu=0,1,2,3$ is implied. We choose $A_{\mu} = \left(-x F, 0, 0, 0\right)$, where $F$ is the strength of the electromagnetic field, and $q = -1$ is coupling to the electromagnetic field for the electron. $\gamma^{\mu}$ are the Dirac matrices obeying Clifford algebra $Cl(1,3)$: $\left\{\gamma^{\mu}, \gamma^{\nu}\right\} = 2\eta^{\mu\nu} I_{4\times4}$, where $\eta^{\mu\nu}$ is the space-time metric with the signature fixed to $\eta^{\mu \nu} = \mathrm{diag}\left(1,-1,-1,-1\right)$ and $I_{4\times4}$ is the $4 \times 4$ unity matrix. We set the Dirac $\gamma$-matrices representation in terms of Pauli matrices, the $\sigma_x$, $\sigma_y$ and $\sigma_z$, and $2 \times 2$ unit matrix $I$ as follows:
\begin{align}
    \gamma^0 &= \begin{pmatrix}
        0 & -\sigma_y\\
        -\sigma_y & 0
    \end{pmatrix}; & \gamma^1 &= \begin{pmatrix}
        0 & i \sigma_z \\
        i \sigma_z & 0
    \end{pmatrix}; & \gamma^2 &= \begin{pmatrix}
        0 & - I \\
        I & 0
    \end{pmatrix}; & \gamma^3 &= \begin{pmatrix}
        0 & -i \sigma_x \\
        -i \sigma_x & 0
    \end{pmatrix} \,.
    \label{eq:GammaMatrices}
\end{align}
The $\bar{\Psi} = \Psi^{\dagger}\gamma^{0}$ in Eq.~(\ref{eq:action}) is a Dirac adjoint. By minimizing the action in Eq.~(\ref{eq:action}), we arrive at the Lagrange-Euler equations of motion for a quantum field~\cite{landau1975classical}. For a given Lagrangian, this result leads to the standard Dirac equation in covariant form: $i \gamma^{\mu} D_{\mu} \Psi = 0$. Multiplying this equation from the right by $\gamma^0$, we transition to the Hamilton form of the Dirac equation in terms of Dirac matrices $\vec{\alpha} = \gamma^0\gamma^{i}$, $i=1,2,3$, and $\beta = \gamma^0$: $i \partial \Psi / \partial t = \mathcal{H} \Psi$, where
\begin{eqnarray}
    \mathcal{H} = \vec{\alpha} \vec{\kappa} - I_{4 \times 4}\, \cdot x F = \begin{pmatrix}
        H_{\mathbf{K}} & 0 \\
        0 & H_{\mathbf{K}^{\prime}}
    \end{pmatrix}\, ,  
\end{eqnarray}
with $\kappa_{x,y,z} = -i \partial_{x,y,z}$.We explicitly write the matrices $\vec{\alpha}$ and $\beta$ as
\begin{align}
    \alpha_x &= \begin{pmatrix}
        \sigma_x & 0 \\
        0 & \sigma_x
    \end{pmatrix}; & \alpha_y &= \begin{pmatrix}
        -\sigma_y & 0 \\
        0 & \sigma_y 
    \end{pmatrix}; & \alpha_z &= \begin{pmatrix}
    \sigma_z & 0 \\
    0 & \sigma_z
    \end{pmatrix};& \beta &= \begin{pmatrix}
        0 & -\sigma_y \\
        -\sigma_y & 0
    \end{pmatrix}\, ,
\end{align}
where they obey the following relations: $\alpha^2_i = \beta^2 = I_{4\times 4};$ $\left\{\alpha_i, \beta\right\} = 0;$ $\left\{\alpha_i, \alpha_j \right\} = \delta_{ij} I_{4\times4}$, for $i,j=1,2,3$.
The $4 \times 4$ Hamiltonian $\mathcal{H}$ features two chiral sectors or valleys denoted as $\mathbf{K}$ and $\mathbf{K}^{\prime}$. If we define $\gamma^5 = -i \gamma^0 \gamma^1 \gamma^2 \gamma^3$, which anti-commutes with all $\gamma$-matrices and squares to unity matrix~\cite{berestetskii1982quantum}:
\begin{equation}
    \gamma^5 = \begin{pmatrix}
        -I & 0 \\
        0 & \phantom{-}I
    \end{pmatrix} \, ,
\end{equation}
then the chiral projector $P = \left(I_{4\times4} \pm \gamma^5\right)/2$ is a valley projector. If the $4$-component bispinor is an eigenvector of the defined chiral operator $\gamma_5$, $\gamma_5 \Psi = \pm \Psi$, then it is also an eigenvector of $H = P \mathcal{H} P^{\dagger}$. Since $H$ is effectively a $2\times 2$ Hamiltonian, the bispinor is essentially a $2$-component spinor $\Psi = \left(\phi_A, \phi_B \right)^\mathrm{T}$, where instead of regular spin up and down components, we use a honeycomb lattice sublattice components A and B [see Fig.~\ref{fig:HoneycombLatticeChiralAnomaly}(a)]. Thus, our initial problem with $4$ equations decays into $2$ independent problems with $2$ equations each. Finally, we confine our massless fermion to $xy$-plane, i.e. $\Psi (x,y,z,t) \equiv \Psi (x,y,t)$ so that $\kappa_z \Psi (x,y,t) = 0$, and impose hard-wall boundary condition on the spinor $\Psi$ along $x$-axis in both chiral sectors, i.e. valleys $\mathbf{K}$ and $\mathbf{K}^{\prime}$:
\begin{equation}
    \left(I -  \sigma_z\right)\Psi \big|_{x=\pm L/2} = 0 \, ,
    \label{eq:BoundaryConditions}
\end{equation}
where $L$ is the size of the stripe which to fermions are confined. Now, fermions here are free to move only perpendicular to the field $F$ along $y$-axis, such that we have $\Psi (x,y,t) = \Psi (x) \exp\left[i \kappa_y y - i \varepsilon t\right]$ and the system turns effectively into $1+1$ space-time.
 
The two $2 \times 2$ time-independent Dirac equations for $\Psi(x)$ read:
\begin{equation}
    \left( - i \sigma_x \partial_x \mp \sigma_y \kappa_y - F x\right) \Psi = \varepsilon \Psi \, ,
    \label{eq:DiracEquation4GNRPauliMatricesEfieldKprime}
\end{equation}
where the upper and lower signs of $\sigma_y \kappa_y$ identify the ${\bf K}$ and $\mathbf{K}^{\prime}$ valleys, respectively. We then examine the expression of the spinors at the $\mathbf{K}$ and $\mathbf{K}^{\prime}$ point. By multiplying both side of Eq.~(\ref{eq:DiracEquation4GNRPauliMatricesEfieldKprime}) by $i \sigma_x$, we get
\begin{equation}
    \left( \partial_x \pm \sigma_z \kappa_y - i \sigma_x (F x + \varepsilon)\right) \Psi = 0 \, .
    \label{eq:DiracEquation4GNRPauliMatricesEfield0}
\end{equation}
with the upper and lower signs of the $\sigma_z \kappa_y$ designate the ${\bf K}$ and $\mathbf{K}^{\prime}$ valleys, respectively. Starting here, we use the upper and lower signs in each equation to denote the $\kappa_y$ sign assignments on the ${\bf K}$ and $\mathbf{K}^{\prime}$ valleys. It is evident that the external field $F$ shows a term proportional to $\sigma_x$, resulting to a sublattice mixing within the system. In terms of the spinor components $\phi_{A,B}$, the set of equations take an explicit form of:
\begin{eqnarray}
    \partial_x \phi_A \pm \kappa_y \phi_A - i \left(\varepsilon + F x\right) \phi_B &=& 0 \,, 
    \label{eq:DiracEquation4GNRPauliMatricesEfield1}\\
    \partial_x \phi_B \mp \kappa_y \phi_B - i \left(\varepsilon + F x\right) \phi_A &=& 0 \,.
    \label{eq:DiracEquation4GNRPauliMatricesEfield2}
\end{eqnarray}
When $\varepsilon + F x = 0$ and $\kappa_y \neq 0$, we get the flat "zero"-energy modes which will be shown later in section \ref{sec:ZEM}. We then consider a new variable $\alpha = \varepsilon + F x$ and let this $\alpha \neq 0$. Starting from Eq.~(\ref{eq:DiracEquation4GNRPauliMatricesEfield2}), we can express $\phi_A$ as $\phi_A = \left(\partial_x \phi_B \mp \kappa_y \phi_B\right)/\left(i \alpha \right)$ and substitute this expression into Eq.~(\ref{eq:DiracEquation4GNRPauliMatricesEfield1}). This yields the following equation for the spinor component $\phi_B$:
\begin{equation}
    \partial_{xx} \phi_B - \frac{F}{\alpha} \partial_x \phi_B \pm \left(\frac{F \kappa_y}{\alpha} + \kappa_y^2 - \alpha^2\right) \phi_B = 0 \, .
    \label{eq:DiracEquation4GNRSingleComponentphiBEfield}
\end{equation}
Next, we introduce a change of the independent variable, $\xi = \alpha / F = \varepsilon / F + x$, into Eq.~(\ref{eq:DiracEquation4GNRSingleComponentphiBEfield}) that results to:
\begin{equation}
    \partial_{\xi \xi} \phi_B - \frac{1}{\xi} \partial_{\xi} \phi_B \pm \left(\frac{\kappa_y}{\xi} \mp \kappa_y^2 \pm F^2 \xi^2\right) \phi_B = 0 \, 
\label{eq:DiracEquation4GNRSingleComponentphiBEfield2}
\end{equation}
with the upper and lower signs for the ${\bf K}$ and $\mathbf{K}^{\prime}$ valleys, respectively. Setting the boundary conditions according to Eq.~(\ref{eq:BoundaryConditions}) yields $\phi_B(\xi_1) = 0$ and $\phi_B(\xi_2) = 0$, where $\xi_1 = (\varepsilon/F) + (L/2)$ and $\xi_2 = (\varepsilon/F) - (L/2)$. 

This Eq.~(\ref{eq:DiracEquation4GNRSingleComponentphiBEfield2}) is a second order differential equation where the singular points can be found at $\xi=0$ and $\xi=\infty$. The nature of these two singularities comprises a regular singular point at $\xi=0$ and an irregular singular point at $\xi=\infty$. Hence, Eq.~(\ref{eq:DiracEquation4GNRSingleComponentphiBEfield2}) resembles a Bessel equation, which represents a special case of a hypergeometric equation with three regular poles, where two regular poles have merged leading to the irregular singularity. We can then employ the power series method~\cite{kreyszig2020advanced} to obtain a solution of Eq.~(\ref{eq:DiracEquation4GNRSingleComponentphiBEfield2}) around the remaining regular singular point $\xi=0$.

The power series method seeks to find a solution to a differential equation using an infinite power series expansion of the form $S=\sum_{n=0}^{\infty} a_n \xi^n$, establishing a recursive relationship between the series coefficients $a_n$’s. Applying directly this method to Eq.~(\ref{eq:DiracEquation4GNRSingleComponentphiBEfield2}) fails due to a wide range of terms leading to a power series with largely shifted degrees of $\xi$’s. To reduce the power series, we can isolate the terms and set the first $4$ coefficients to zero in the final series. Unfortunately, this process disrupts the the recursive relation between $a_n$ coefficients.

To address this issue, we aim to simplify the components in the Eq.~(\ref{eq:DiracEquation4GNRSingleComponentphiBEfield2}) by eliminating terms within $\left(\left(\kappa_y/\xi\right) \mp \kappa_y^2 \pm F^2 \xi^2\right) \phi_B$. We can take a function $\phi_B (\xi) = u(\xi) \exp\left(\pm \kappa_y \xi\right)$ to remove the terms $\kappa_y^2$ and $\kappa_y/\xi$, where ``$+$'' is for the $\mathbf{K}$ valley while ``$-$'' is for the $\mathbf{K}^{\prime}$ valley. From a physical standpoint, this substitution can be interpreted as a limiting solution under a zero external field $F=0$ or an equivalent large $\kappa_y$ limit for $F\neq 0$. This substitution results in:
\begin{equation}
    \partial_{\xi \xi} u \pm \left(2 \kappa_y \mp \frac{1}{\xi} \right) \partial_{\xi} u + F^2 \xi^2 u = 0 \, .
    \label{eq:DiracEquation4GNRSingleComponentphiBEfield2uSub}
\end{equation}
The presence of the term $F^2 \xi^2 u$ leads to a power series significantly far from the other terms due to the $\xi^2$ factor. We can fix this issue by introducing another substitution $u(\xi) = g(\xi) \exp\left(\pm i F \xi^2 / 2 \right)$, which resembles  the solution derived for the case $\kappa_y = 0$ corresponding to the Eq.~(\ref{eq:DiracEquation4GNRinEfieldComponentphiBinKSol}) later in section \ref{sec:ZEM}, and where ``$+$'' is for the $\mathbf{K}$ valley while ``$-$'' is for the $\mathbf{K}^{\prime}$ valley. The resulting equation then becomes:
\begin{equation}
    \partial_{\xi \xi} g \pm \left(2 i F \xi + 2 \kappa_y \mp \frac{1}{\xi} \right) \partial_{\xi} g + 2 i \kappa_y F \xi  g = 0 \, .
\label{eq:DiracEquation4GNRSingleComponentphiBEfield2ugSub}
\end{equation}
To simplify Eq.~(\ref{eq:DiracEquation4GNRSingleComponentphiBEfield2ugSub}), we introduce new parameters: $\mu=2\kappa_{y}$, $\nu=2iF$, and $\lambda=2i\kappa_{y}F$ which yields to a modified equation:
\begin{equation}
    \partial_{\xi \xi} g \pm \left(\nu \xi + \mu \mp \frac{1} {\xi} \right) \partial_{\xi} g + \lambda \xi g=0 .
\label{eq:DiracEquation4GNRSingleComponentphiBEfield2ugSimplify}
\end{equation}
We emphasize that in Eq.~(\ref{eq:DiracEquation4GNRSingleComponentphiBEfield2ugSimplify}), the condition $F\neq0$ guarantees that both $\nu$ and $\lambda$ are non-zero.

The series expansion using the power series method becomes applicable now. By defining $g(\xi) \equiv S = \sum_{n=0}^{\infty} a_n \xi^n$, we obtain:
\begin{equation}
    \sum_{n=0}^{\infty} a_n n (n-1) \xi^{n-2} \pm \left( \nu \xi \sum_{n=0}^{\infty} a_n n \xi^{n-1} + \mu \sum_{n=0}^{\infty} a_n n \xi^{n-1} \mp \sum_{n=0}^{\infty} a_n n \xi^{n-2} \right) + \lambda \sum_{n=0}^{\infty} a_n \xi^{n+1} = 0 \, .
\label{eq:DiracEquation4GNRSingleComponentphiBEfield2ugSubPE}
\end{equation}
After several derivations, we modify the dummy summation variable $n$ to $k$ in each power series and convert the $\xi$ exponents into new summation variables:
\begin{equation}
    -\frac{a_{1}}{\xi} \pm \mu a_{1} + \sum_{k=1}^{\infty}  \left[a_{k+2} \left(k + 2 \right) k \pm \nu a_{k}k \pm \mu a_{k+1} \left(k + 1 \right) + \lambda a_{k-1} \right] \xi^{k} =0 \, ,
\label{eq:DiracEquation4GNRSingleComponentphiBEfield2ugSubPE1}
\end{equation}
where we gathered the non-zero terms and grouped all summations starting from $k=1$. It becomes clear that the equation is satisfied when we set $a_1 \equiv 0$, and all the terms within the square bracket $\left[ \ldots \right]$ sum up to zero. This condition implies the following recurrence relation between the coefficients $a_{k+2}$:
\begin{equation}
    a_{k+2} = \mp \frac{\mu \left(k + 1 \right)}{k \left(k + 2 \right)} a_{k+1} \mp \frac{\nu} {k+2} a_{k}-\frac{\lambda} {k \left(k + 2\right)} a_{k-1} \, .
\label{eq:DiracEquation4GNRSingleComponentphiBEfieldRR}
\end{equation}
Alternatively, we can revert to using the summation variable $n$ that leads to the following recurrence relation:
\begin{equation}
    a_{n}=\mp\frac{\mu\left(n-1\right)}{n\left(n-2\right)}a_{n-1}\mp\frac{\nu}{n}a_{n-2}-\frac{\lambda}{n\left(n-2\right)}a_{n-3} \, , \label{eq:DiracEquation4GNRSingleComponentphiBEfieldRR1}
\end{equation}
where $n=3, 4, \ldots$, and the upper and lower signs correspond again to the $\mathbf{K}$ and $\mathbf{K}^{\prime}$ valleys, respectively. Since $a_1 \equiv 0$, we have only two unknown coefficients remaining, namely, $a_0$ and $a_2$. 

By separately collecting the power series terms involving $a_0$ and $a_2$, we can divide the entire series into two distinct special functions represented by $R(\xi, \pm \mu, \pm \nu, \lambda)$ and $V(\xi, \pm \mu, \pm \nu, \lambda)$ similar to how the Airy functions~\cite{griffiths2018introduction} serve as solutions to the triangular potential well. However, these special functions $R$ and $V$ nor their derivatives are not Airy functions as we shall see later. Each of these $R$ and $V$ functions is designated by a power series with coefficients determined from the recursive relation in Eq.~(\ref{eq:DiracEquation4GNRSingleComponentphiBEfieldRR1}) and from two specific initial conditions. Specifically, we set $a_0=1$ and $a_2=0$ for the $R$-function, while for the $V$-function, we set $a_0=0$ and $a_2 = 1$. The general solution of Eq.~(\ref{eq:DiracEquation4GNRSingleComponentphiBEfield2ugSub}) can then be expressed in the form:
\begin{equation}
    g(\xi) = C_1 R(\xi, \pm \mu, \pm \nu, \lambda) + C_2 V(\xi, \pm \mu, \pm \nu, \lambda) \, ,
\label{eq:DiracEquation4GNRSingleComponentphiBEfieldGenSol}
\end{equation}
where $C_{1,2}$ represent the arbitrary constants to be determined from the boundary conditions. Essentially, these constants indicate the free parameters $a_0$ and $a_2$.

Now, we can write down the general solution for Eq.~(\ref{eq:DiracEquation4GNRSingleComponentphiBEfield2}) explicitly as:
\begin{equation}
    \phi_B(\xi) = \exp\left(\pm \kappa_y \xi \right) \exp\left(\pm \frac{i F \xi^2}{2} \right) \left[C_1 R(\xi, \pm \mu, \pm \nu, \lambda) + C_2 V(\xi, \pm \mu, \pm \nu, \lambda)\right] \, ,
\label{eq:DiracEquation4GNRSingleComponentphiBEfieldGenSol1}
\end{equation}
where the upper ``$+$'' and lower ``$-$'' signs correspond to the ${\bf K}$ and $\mathbf{K}^{\prime}$ valleys, respectively. Applying the boundary conditions from  Eq.~(\ref{eq:DiracEquation4GNRSingleComponentphiBEfield2}) where $\phi_B(\xi_1) = \phi_B(\xi_2) = 0$ and $\xi_{1,2} = (\varepsilon/F) \pm (L/2)$, we get the following relations of coefficients $C_1$ and $C_2$: 
\begin{equation}
    C_{1}=-C_{2}\frac{V\left(\xi_{1},\pm\mu,\pm\nu,\lambda\right)}{R\left(\xi_{1},\pm\mu,\pm\nu,\lambda\right)} \, ,
    \label{eq:DiracEquation4GNRSingleComponentphiBEfieldCrelation}
\end{equation}
\begin{equation}
    C_{1}=-C_{2}\frac{V\left(\xi_{2},\pm\mu,\pm\nu,\lambda\right)}{R\left(\xi_{2},\pm\mu,\pm\nu,\lambda\right)} \, .
    \label{eq:DiracEquation4GNRSingleComponentphiBEfieldCrelation1}
\end{equation}
Substituting this coefficient relation in Eq.~(\ref{eq:DiracEquation4GNRSingleComponentphiBEfieldCrelation}) gives us a general equation of $\phi_ B$:
\begin{equation}
    \phi_{B}=C_{2}\exp\left(\pm\kappa_{y}\xi\right)\exp\left(\frac{\pm iF\xi^{2}}{2}\right)M\left(\xi\right)
    \label{eq:DiracEquation4GNRSingleComponentphiBEfieldGenSol2}
\end{equation}
where $M\left(\xi\right)$ is defined in terms of the special functions $R$ and $V$:
\begin{equation}
M\left(\xi\right) \equiv \left[R \left(\xi_{1}, \pm \mu, \pm \nu, \lambda\right) V \left( \xi, \pm \mu, \pm \nu, \lambda \right) - R \left( \xi, \pm \mu, \pm \nu, \lambda \right) V \left (\xi_{1}, \pm \mu, \pm \nu, \lambda \right) \right] /R \left( \xi_{1}, \pm \mu, \pm \nu, \lambda \right)
\label{eq:DiracEquation4GNRSingleComponentphiBEfieldGenSol2-1}
\end{equation}
As for the spinor component $\phi_A$, we recall that $\phi_{A}=\frac{1}{i \xi F}\left(\partial_{x}\phi_{B}\mp\kappa_{y}\phi_{B}\right)$ with the upper ``$-$'' and lower ``$+$'' signs indicating the ${\bf K}$ and $\mathbf{K}^{\prime}$ valleys, respectively. The spinor $\phi_A$-component expressed in terms of special functions $R$ and $V$ is then:
\begin{equation}
    \phi_{A} = \frac{C_{2}}{i F \xi} \exp \left(\pm \kappa_{y} \xi \right) \exp \left( \pm \frac{i F \xi^{2}}{2} \right) \left[ \partial_{\xi} \pm i F \xi \right] M \left (\xi \right)
\label{eq:DiracEquation4GNRSingleComponentphiAEfieldGenSol}
\end{equation}
where the upper ``$+$'' and lower ``$-$'' signs correspond to the ${\bf K}$ and $\mathbf{K}^{\prime}$ valleys, respectively. Combining Eqs.~(\ref{eq:DiracEquation4GNRSingleComponentphiBEfieldGenSol2}) and (\ref{eq:DiracEquation4GNRSingleComponentphiAEfieldGenSol}), and then normalizing the full expression yield the normalized spinor $\Psi$ solution represented by:
\begin{equation}
    \Psi =  \begin{pmatrix}
    \phi_A \cr
    \phi_B
    \end{pmatrix} = 
    C \exp \left(\pm \kappa_{y} \xi \right) \exp \left( \pm \frac{i F \xi^{2}}{2} \right) \exp \left( i \kappa_{y} y \right)
    \begin{pmatrix}
    \frac{1}{i F \xi} \partial_{\xi} M \left (\xi \right) \pm M \left (\xi \right) \cr
    M \left (\xi \right)\end{pmatrix}\ .
    \label{eq:DiracEquation4GNRSingleComponentPsiSol}
\end{equation}
where the normalizing constant $C$ is given by:
\begin{equation}
    C = \left(\sqrt {\int_ {\frac{\varepsilon}{F}-\frac{L}{2}} ^ {\frac{\varepsilon}{F} + \frac{L}{2}} \exp \left(\pm2 \kappa_{y} \xi \right) \left[ \left| \frac{1}{i F \xi} \partial_{\xi} M \left( \xi \right) \pm M \left( \xi \right) \right| ^{2} + \left| M \left( \xi \right) \right| ^{2} \right]} \right) ^{-1}
\label{eq:DiracEquation4GNRSingleComponentPsiSol-1}
\end{equation}and the function $M\left( \xi \right)$ is defined in Eq.~(\ref{eq:DiracEquation4GNRSingleComponentphiBEfieldGenSol2-1}) with the special functions $R$ and $V$, and the upper ``$+$'' and lower ``$-$'' signs correspond to the ${\bf K}$ and $\mathbf{K}^{\prime}$ valleys, respectively. It is important to note here that
\begin{equation}
    \phi_{B} \left (\frac{\epsilon}{F} + \frac{L}{2} \right) = \phi_{B} \left(\frac{\epsilon}{F} - \frac{L}{2} \right) = 0 \iff M \left(\xi_{1} \right) = M\left (\xi_{2}\right) = 0\ .
    \label{eq:DiracEquation4GNRSingleComponentPsiSol-2}
\end{equation}
The homogeneous system of linear equations given by Eqs.~(\ref{eq:DiracEquation4GNRSingleComponentphiBEfieldCrelation}) and (\ref{eq:DiracEquation4GNRSingleComponentphiBEfieldCrelation1}) has non-trivial solutions only when the determinant of its matrix equals zero, which then leads to a dispersion equation: 
\begin{equation}
    \exp\left(\frac{iF\left(\xi_{1}^{2}+\xi_{2}^{2}\right)}{2}\right) M\left(\xi_{2}\right) R \left( \xi_{1}, \pm \mu, \pm \nu, \lambda \right) = 0\, .
    \label{eq:DiracEquation4GNRDispersionRelation}
\end{equation}
For a given $F$ and $L$, Eq.~(\ref{eq:DiracEquation4GNRDispersionRelation}) establishes the connection between $\varepsilon$ and $\kappa_y$. One can retain the complex exponent in Eq.~(\ref{eq:DiracEquation4GNRDispersionRelation}), since it serves as a regularizing factor which ensures that the zeros of the dispersion equation are determined solely by its real part.

\section{Discussion}
In this section, we shall investigate some limiting cases of the above-described problem for the Dirac equation with the boundary conditions~(\ref{eq:BoundaryConditions}). The zero-energy modes (ZEMs) of the Dirac fermions shall be explicitly shown with an single $\kappa_y$-point solution to the two-component spinor $\Psi$ function. Finally, we shall present the unique properties of $R$ and $V$ special functions in a form of complex plots and distinguish these two functions from the well-known Airy functions.

\subsection{\label{sec:Spinor}Bulk modes}
 
The Eqs.~(\ref{eq:DiracEquation4GNRPauliMatricesEfield1}) and~(\ref{eq:DiracEquation4GNRPauliMatricesEfield2}) in the case $F=0$ reads
\begin{eqnarray}
    - i \partial_x \phi_B \pm i \kappa_y \phi_B &=& \varepsilon \phi_A \, , \label{eq:DiracEquation4GNR1} \\
    - i\partial_x \phi_A \mp i \kappa_y \phi_A &=& \varepsilon \phi_B \, , \label{eq:DiracEquation4GNR2} 
\end{eqnarray}
where $\kappa_y$ is the electron momentum along the ribbon measured with respect to the Dirac point and the upper and lower signs of $\sigma_y \kappa_y$ identify the ${\bf K}$ and $\mathbf{K}^{\prime}$ valleys, respectively. We want to find the secular equations from these two set of differential equations. Expressing $\phi_A$ from Eq.~(\ref{eq:DiracEquation4GNR1}) as $\phi_A = - \left(i/\varepsilon\right)\partial_x \phi_B \pm \left(i \kappa_y / \varepsilon \right) \phi_B$ and putting it into Eq.~(\ref{eq:DiracEquation4GNR2}), we arrive at a single second order equation:
\begin{equation}
    - \partial_{xx} \phi_B + \kappa_y^2 \phi_B = \varepsilon^2 \phi_B \, .
    \label{eq:DiracEquation4GNR2ndOrder} 
\end{equation}
The general solution of Eq.~(\ref{eq:DiracEquation4GNR2ndOrder}) is  $\phi_B = C_1 e^{zx} + C_2 e^{-zx}$, with $z = \sqrt{\kappa_y^2 - \varepsilon^2}$ and $C_{1,2}$ as constants determined from the boundary and normalization conditions. We use the boundary conditions~(\ref{eq:BoundaryConditions}):
\begin{equation}\phi_B\left(-\frac{L}{2}\right) = \phi_B\left(+\frac{L}{2}\right) = 0
\label{eq:DiracEquationhbGNRbc}\end{equation}
which yield the following set of equations:
\begin{eqnarray}
    C_1 e^{z \cdot (-L/2)} + C_2 e^{-z \cdot (-L/2)} &=& 0\, , \nonumber \\ C_1 &=& - C_2 e^{zL} \,, \label{eq:DiracEquationBC1} \\
    C_1 e^{z L/2} + C_2 e^{-z L/2} &=& 0 \, , \nonumber \\ 
    2 e^{-z L/2} C_1 \sinh \left(z L\right)&=& 0 \, . \label{eq:DiracEquationBC2}
\end{eqnarray}
Eq.~(\ref{eq:DiracEquationBC2}) shows that the secular equation $\sinh \left(zL\right) = 0$ leads to $\varepsilon = \pm \left|\kappa_y\right|$. However, this solution is spurious since then $\phi_B$ is identically vanishing because of $z=0$. The correction description around zero-energy is shown in the next section. 

Another approach is to substitute $z = i \kappa_x$ that leads to quantization of the transverse electron momentum $\kappa_x$: $\kappa_x = \pi n / L$, with $n$ being integer. The valence and conduction energy bands are described by $\varepsilon = \pm \sqrt{\kappa_y^2 + \left(\pi n / L\right)^2}$, where $n =  \pm 1, \pm 2, \ldots$. Then Eqs.~(\ref{eq:DiracEquationhbGNRbc}) take form:
\begin{eqnarray}
    C_1 e^{- i \kappa_x L/2} + C_2 e^{i \kappa_x L/2} &=& 0 \,, \nonumber \\
    C_1 e^{i \kappa_x L/2} + C_2 e^{- i \kappa_x L/2} &=& 0 \, . \label{eq:DiracEquationbc2}
\end{eqnarray}
These homogeneous set of linear equations from Eq.~(\ref{eq:DiracEquationbc2}) can have non-trivial solutions when the determinant is zero, which defines the quantization of $\kappa_x$. From the second equation in Eqs.~(\ref{eq:DiracEquationbc2}), we get $C_2 = - C_1 e^{i \kappa_x L}$ so that $\phi_B = - 2 i C_1 e^{i \kappa_x L/2} \sin \left[\kappa_x \left(L/2 - x\right)\right]$ and $ \phi_A = \pm 2 C_1 e^{i \kappa_x L/2} \cos\left[\kappa_x \left(L/2 - x\right) + \varphi \right]$, where `$+$' and `$-$' in $\phi_A$ are used for the bands above and below $\varepsilon=0$, respectively, and $\varphi = \arcsin \left( \kappa_y/\sqrt{\kappa_y^2+\kappa_x^2}\right)$. Thus, the normalized spinor is:
\begin{equation}
    \Psi = \begin{pmatrix}
    \phi_A \cr
    \phi_B
    \end{pmatrix} =  \frac{1}{\sqrt{L}} \begin{pmatrix}
    \pm \cos \left[\kappa_x \left(L/2 - x \right) + \varphi \right]    \cr
    -i \sin \left[\kappa_x \left(L/2 - x \right) \right]
    \end{pmatrix}  \, ,
    \label{eq:DiracEquation4GNRSpinorPsi}
\end{equation}
where the global phase $e^{i \kappa_x L/2}$ is ignored. It is seen from Eq.~(\ref{eq:DiracEquation4GNRSpinorPsi}) that the obtained spinor describes extended or bulk states.

\subsection{\label{sec:ZEM}Zero-energy edge mode}

The zero-energy modes (ZEMs) of Dirac fermions have been interesting subject in quantum field theory since its first introduction by Jackiw and Rebbi~\cite{jackiw1976solitons} in 1976. Currently, ZEMs play pivotal role in topological phases of condensed matter~\cite{shen2012topological}. We encounter a ZEM by considering Eq.~(\ref{eq:DiracEquation4GNRPauliMatricesEfieldKprime}) in the ${\bf K}$-valley for $F=0$:
\begin{equation}
    \left( - i \sigma_x \partial_x - \sigma_y \kappa_y \right) \Psi = \varepsilon \Psi \, ,
    \label{eq:DiracEquation4GNRPauliMatrices}
\end{equation}
where $\Psi = \left(\phi_A, \phi_B \right)^\mathrm{T}$. We set $\varepsilon = 0$ and multiply both sides by $\sigma_x$ to find:
\begin{equation}
    \partial_x \Psi = - \sigma_z \kappa_y \Psi \, .
    \label{eq:DiracEquation4GNRPauliMatricesZEM}
\end{equation}
If the spinor $\Psi$ is an eigenvector of $\sigma_z$, that is, $\sigma_z V_{\lambda} = \lambda V_{\lambda}$, then the Eq.~(\ref{eq:DiracEquation4GNRPauliMatricesZEM}) above reduces to a scalar equation where for $\lambda_{+}=1, V_{+} = \left(1, 0\right)^{\mathrm{T}}$ and for $\lambda_{-}=-1, V_{-} = \left(0, 1\right)^{\mathrm{T}}$.Then, the $\Psi_{\lambda} = V_{\lambda} \, \chi (x)$, where the scalar function $\chi(x)$ is found from:
\begin{equation}
    \partial_x \chi (x) = - \lambda \kappa_y \chi(x) \, .
    \label{eq:DiracEquation4GNRZEMscalar}
\end{equation}
From the solution $\chi_{\pm}(x) = A \exp\left(-\lambda_{\pm} \kappa_y x\right)$, the spinors can be described with:
\begin{equation}
    \Psi_{\pm}(x) = V_{\pm} A \exp\left(- \lambda_{\pm} \kappa_y x\right)\, ,
    \label{eq:DiracEquation4GNRPauliMatricesZEMSolution}
\end{equation}
where they exponentially change for $\kappa_y \neq 0$. These spinors are similar to those combined to construct a normalizible Jackiw-Rebbi soliton~\cite{jackiw1976solitons} satisfying the boundary conditions $\Psi_{\pm}(x\rightarrow \pm \infty) \rightarrow \mathbf{0}$. Our boundary conditions given in Eq.~(\ref{eq:BoundaryConditions}) are different. Therefore, we need to pick out only one spinor out of two provided by Eq. (\ref{eq:DiracEquation4GNRPauliMatricesZEMSolution}), the one that satisfies the boundary conditions in Eq.~(\ref{eq:BoundaryConditions}). Here we do not need a combination of spinors in order to produce a normalizable solution since our system has finite size along $x$-axis. Considering the finite size of the system, the counterpart of the Jackiw-Rebbi soliton emerges as a fully sublattice polarized mode. Evidently, the arbitrary constant in the general solution of the first-order Eq.~(\ref{eq:DiracEquation4GNRZEMscalar}) corresponds solely to the normalization degree of freedom. Thus, the appropriate solution is
\begin{equation}
    \Psi_{+}(x) = \begin{pmatrix}
    1 \cr
    0
    \end{pmatrix} C_1 \exp\left(-\kappa_y x\right) \, .
    \label{eq:DiracEuationhbGNRJackiwRebbiExponentialGrowth}
\end{equation}
We can find the normalization constant as follows:
\begin{eqnarray}
    \int_{-L/2}^{L/2} \Psi_{+}^{\dagger}(x) \Psi_{+}(x) d\, x &=& 1 \, , \nonumber \\
    \left|C_1\right|^2 \int_{-L/2}^{L/2} \exp\left(-2 \kappa_y x\right) d\, x &=& 1\, , \nonumber \\
    \left|C_1\right| &=& \sqrt{\frac{\kappa_y}{\sinh (\kappa_y L)}} \, ,
\end{eqnarray}
which allows us to write down the ZEM normalized spinor:
\begin{equation}
    \Psi_{+}(x) = \begin{pmatrix}
    1 \cr
    0\end{pmatrix} \sqrt{\frac{\kappa_y}{\sinh (\kappa_y L)}}\, \exp\left(-\kappa_y x\right) \, .
    \label{eq:DiracEuationhbGNRJackiwRebbiExponentialGrowthNorm}
\end{equation}
Varying $\kappa_y$ around the Dirac point ($\kappa_y = 0$) shifts the ZEM wave function from one edge to another while preserving sublattice polarization and maintaining its chiral nature. The ZEM wave function loses its localization only at the proximity of the Dirac point. The wave function in Eq.~(\ref{eq:DiracEuationhbGNRJackiwRebbiExponentialGrowthNorm}) is an eigenfunction of the $2 \times 2$ chiral symmetry operator $\mathcal{S} = \sigma_z$. 

The second spinor in Eq.~(\ref{eq:DiracEquation4GNRPauliMatricesZEMSolution}) satisfies boundary conditions $\left(I +  \sigma_z\right)\Psi \big|_{x=\pm L/2} = 0$ [cf. Eq.~(\ref{eq:BoundaryConditions})]. Physically, this means imposing condition on $A$ sublattice. The $A$ sublattice counterpart of Eq.~(\ref{eq:DiracEuationhbGNRJackiwRebbiExponentialGrowthNorm}) reads:
\begin{equation}
    \Psi_{-}(x) = \begin{pmatrix}
    0 \cr
    1\end{pmatrix} \sqrt{\frac{\kappa_y}{\sinh (\kappa_y L)}}\, \exp\left(\kappa_y x\right) \, .
    \label{eq:DiracEuationhbGNRJackiwRebbiExponentialGrowthNormA}
\end{equation}
In the $2 \times 2$ Dirac equation $i\partial_t\Psi = H \Psi$, where $H$ is defined by Eq.~(\ref{eq:DiracEquation4GNRPauliMatrices}), the hole (or charge conjugation) symmetry operator is denoted as $\Xi=\sigma_x K$ with $K$ representing the the complex conjugation. Upon applying the $\Xi$ operator to $\Psi_{\pm}$ as described in Eqs.~(\ref{eq:DiracEuationhbGNRJackiwRebbiExponentialGrowthNorm}) and~ (\ref{eq:DiracEuationhbGNRJackiwRebbiExponentialGrowthNormA}), it is evident that $\Xi \Psi_{\pm} \neq \Psi_{\pm}$ which indicates that the identified ZEM does not conform to Majorana fermion properties. The solutions in the $\mathbf{K}^{\prime}$ valley are essentially the same and can be obtained from those above via substitution: $\kappa_y \rightarrow - \kappa_y$.

\subsection{\label{sec:ZEM in E field}Field independent $\kappa_y=0$ states ladder and $1+1$ chiral anomaly structure}
 
Equation~(\ref{eq:DiracEquation4GNRSingleComponentphiBEfield2}) has a much simpler analytic solution at the Dirac point, $\kappa_y = 0$, where we obtain
\begin{equation}
    \partial_{\xi \xi} \phi_B - \frac{1}{\xi} \partial_{\xi} \phi_B + F^2 \xi^2 \phi_B = 0 \, ,
    \label{eq:DiracEquation4GNRSingleComponentphiBEfieldinK}
\end{equation}
for both $\mathbf{K}$ and $\mathbf{K}^{\prime}$, resulting in the solution space described by:
\begin{equation}
    \phi_B (\xi) = C_1 \cos \left(\frac{F\xi^2}{2}\right) + C_2 \sin \left(\frac{F\xi^2}{2}\right) \, .
    \label{eq:DiracEquation4GNRinEfieldComponentphiBinKSol}
\end{equation}
Such simple solution allows to find analytical expressions for energy levels, which form an equidistant ladder that is independent of the field $F$. The two boundary conditions from Eq.~(\ref{eq:BoundaryConditions}) together with Eq.~(\ref{eq:DiracEquation4GNRinEfieldComponentphiBinKSol}) lead to a set of simultaneous equations:
\begin{eqnarray}
     C_1 \cos \left(\frac{F\xi_1^2}{2}\right) + C_2 \sin \left(\frac{F\xi_1^2}{2}\right) &=& 0 \, , \nonumber \\
     C_1 \cos \left(\frac{F\xi_2^2}{2}\right) + C_2 \sin \left(\frac{F\xi_2^2}{2}\right) &=& 0 \, .
     \label{eq:DiracEquation4GNRinEfieldComponentphiBinKSolBC}
\end{eqnarray}
The set of Eqs.~(\ref{eq:DiracEquation4GNRinEfieldComponentphiBinKSolBC}) has non-zero solutions for coefficients $C_1$ and~$C_2$ only if the determinant of the set matrix is zero. This condition then defines the dispersion equation:
\begin{eqnarray}
      \cos \left(\frac{F\xi_1^2}{2}\right) \sin \left(\frac{F\xi_2^2}{2}\right) - \sin \left(\frac{F\xi_1^2}{2}\right) \cos \left(\frac{F\xi_2^2}{2}\right)  &=& 0 \, , \nonumber \\
     \sin \left[\frac{F}{2}\left(\xi_1^2 - \xi_2^2\right)\right] &=& 0 \,  , \nonumber \\
     \sin\left(\varepsilon L\right) &=& 0 \, ,   \label{eq:DiracEquation4GNRinEfieldComponentphiBinKSolDispersionEquation}
\end{eqnarray}
where we have substituted the definitions of $\xi_{1}$ and $\xi_2$ in the last line. Solving Eq.~(\ref{eq:DiracEquation4GNRinEfieldComponentphiBinKSolDispersionEquation}), a set of field-independent energy states $\varepsilon = \pi n/ L$ is found for $n = 0, \pm 1, \pm 2, \ldots$, which includes also a zero-energy state. 

The state pinned at zero energy for any value of $F$ together with the note made after Eqs.~(\ref{eq:DiracEquation4GNRPauliMatricesEfield1}) and~(\ref{eq:DiracEquation4GNRPauliMatricesEfield2}) reveals the $1+1$ chiral anomaly structure of the energy levels upon $F\neq 0$. Indeed, when $|\kappa_y| \rightarrow \infty$, we have $\alpha(x) = \varepsilon + F x = 0$, so that Eqs.~(\ref{eq:DiracEquation4GNRPauliMatricesEfield1}) and~(\ref{eq:DiracEquation4GNRPauliMatricesEfield2}) 
become of the same type as those in section \ref{sec:ZEM}. Namely, they result in the exponentially localized edge states, therefore we can set $x=\pm L/2$ in $\alpha(x)$. Hence, in general $\varepsilon = \pm FL/2$. Recall, however, that $\mathbf{K}$ and $\mathbf{K}^{\prime}$-valley spinors are localized at opposite edges, thus $\varepsilon = \pm FL/2$ in  $\mathbf{K}$-valley and $\varepsilon = \mp FL/2$ in $\mathbf{K}^{\prime}$-valley in the case $|\kappa_y| \rightarrow \infty$. To summarize, as $\kappa_y$ changes from $-\infty$ to $\infty$, the energy band labeled with $n=0$ changes from $\varepsilon = FL/2$ to $-FL/2$ passing through zero when $\kappa_y = 0$. The picture is reversed in the $\mathbf{K}^{\prime}$-valley. This behaviour of $n=0$ energy bands exhibiting a gauge field dependent chiral anomaly structure is schematically shown in Fig.~\ref{fig:HoneycombLatticeChiralAnomaly}(b).

\subsection{\label{sec:RandVfunctions}Functions $R$ and $V$}
\begin{figure}
\includegraphics[width=0.98\textwidth]{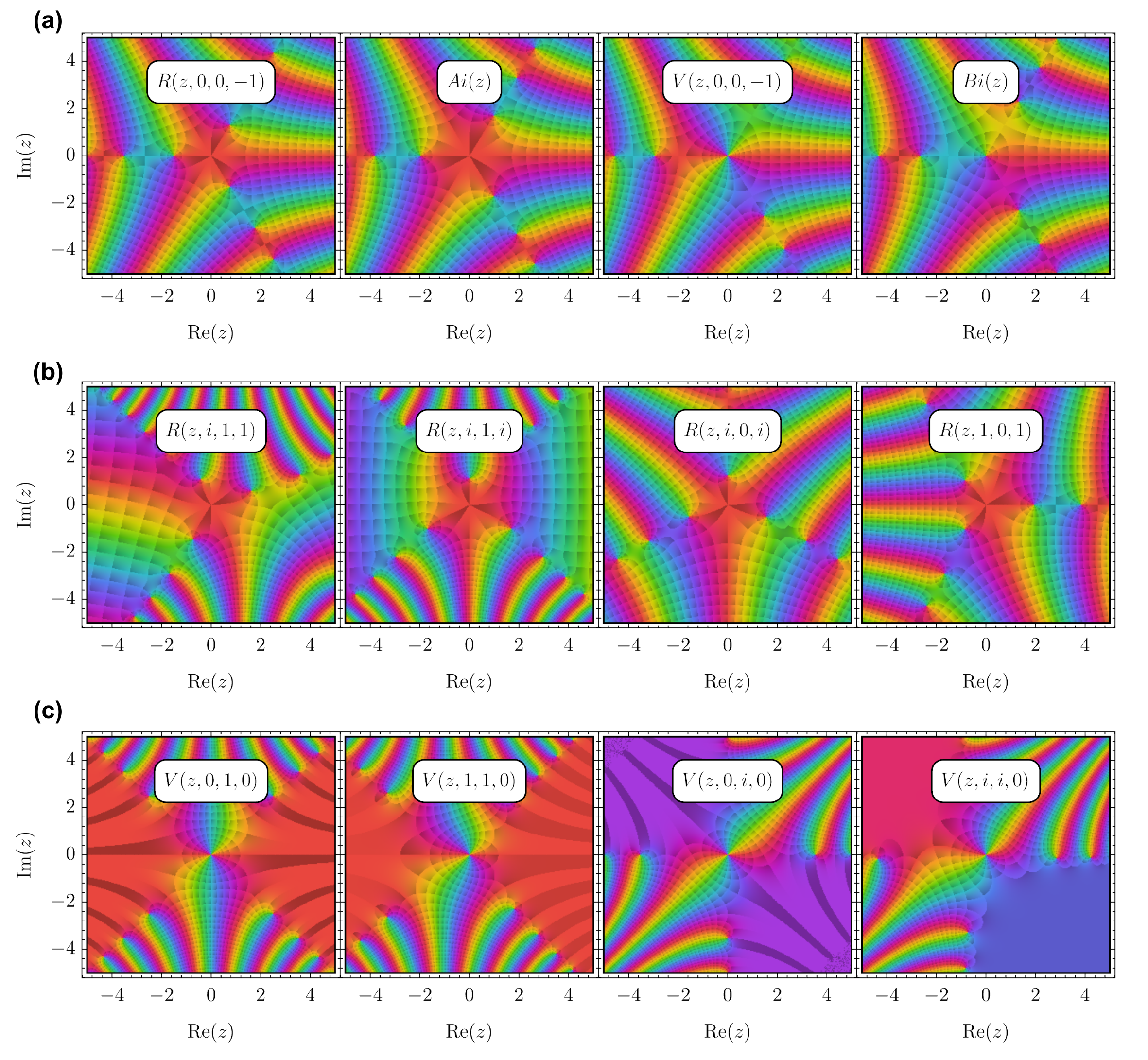}%
\caption{\label{fig:RVAiBi} (a) A comparison between $R$ and $V$ functions and Airy functions of the first $Ai(z)$ and second $Bi(z)$ kind. Multiple zeros and phase patterns are distributed following $C_3$ symmetry, i.e. three rays starting at the origin and forming $2\pi/3$ angles between them. (b) Complex plots of $R$ function for selected values of the inherent parameters $\mu$, $\nu$ and $\lambda$. (c) Same as (b) but for $V$ function. In all plots, the cyclic color function is used for the argument of the complex number ranging from $-\pi$ to $\pi$ represented by hue color change from blue to green via red for a zero phase ("CyclicLogAbsArg" option as defined in Wolfram Mathematica); The color function goes from $-\pi$ (blue) to $0$(red) to $\pi$(green) counterclockwise around zeros, and clockwise around poles.}
\end{figure}

To distinguish functions $R$ and $V$ from the Airy functions, we first examine the recurrence relations given by Eq.~(\ref{eq:DiracEquation4GNRSingleComponentphiBEfieldRR1}) compared to those of the Airy functions. From the Airy equation $y^{\prime\prime} - z y = 0$, the recurrence relations are 
\begin{equation}
    a_{n+2} = \frac{a_{n-1}}{(n+2)(n+1)}\, ,
    \label{eq:RecurrenceRelationsAiryFunctions}
\end{equation}
where $a_2 \equiv 0$. The closest we can get to it is by setting in Eq.~(\ref{eq:DiracEquation4GNRSingleComponentphiBEfieldRR}) $\mu = \nu = 0$ and $\lambda = -1$:
\begin{equation}
    a_{k+2} = \frac{a_{k-1}}{(k+2)k}\, ,
    \label{eq:RecurrenceRelationsRVsimpl}
\end{equation}
but even in this case we have $a_1 \equiv 0$ and denominator $(k+2)k$ instead of $(n+2)(n+1)$ from the recurrence relations of the Airy functions. In Fig.~\ref{fig:RVAiBi}(a), we compare $V(z,0,0,-1)$ function with the Airy function of the second kind $Bi(z)$ by plotting their complex plots. It is clearly seen that while zeros represented by hue vortexes are positioned differently, they follow the same $C_3$ symmetry within the complex plane. Similar regularity is observed for $R$ function and the Airy function of the first kind $Ai(z)$, when $\mu=\nu=0$ and $\lambda = -1$.

Explicitly showing the first few terms in the power series representation (not to be mixed with the Taylor expansion) of Airy functions and $R$ and $V$:
\begin{align}
    Ai(z) &= 1 + \frac{z^3}{(3 \cdot 2)} + \frac{z^6}{(6 \cdot 5) (3 \cdot 2)} + \ldots \, , \label{eq:OurAiryAi}\\
    Bi(z) &= z + \frac{z^4}{(4 \cdot 3)} + \frac{z^7}{(7 \cdot 6) (4 \cdot 3)} + \ldots \, , \label{eq:OurAiryBi}\\
    R(z,\mu,\nu,\lambda) &= 1 - \frac{\lambda  z^3}{3} + \frac{1}{8} \lambda  \mu  z^4 + \frac{\lambda }{15}\left(\nu - \frac{\mu ^2}{2}\right) z^5 + \ldots \, , \\
    V(z, \mu, \nu, \lambda) &= z^2 - \frac{2 \mu  z^3}{3} + 
   \frac{1}{4}\left(\mu^2 - \nu \right) z^4 + \frac{1}{15}\left(-\lambda - 4 \mu  \left(\frac{\mu ^2}{4}-\frac{\nu }{4}\right) + 2 \mu  \nu \right) z^5 + \ldots \, ,
\end{align}
we can see some similarities and differences between our $R$ and $V$ functions and the Airy functions. The power series of $R$ function has a missing $z^2$ term, while $V$ function is missing $1$ term. These are linearly independent monomials. The power series of $Ai(z)$ is missing $z$ term, while $Bi(z)$ is missing $1$ term. Globally, Airy functions are missing $z^2$ terms and all dependent ones as follows from Eq.~(\ref{eq:RecurrenceRelationsAiryFunctions}): $0=a_2=a_5=a_8=\dots = a_{3n+2}$. In contrast, $R$ and $V$-functions are generally missing $z$ terms in their power series representation. When $\mu=\nu=0$, it follows from Eq.~(\ref{eq:RecurrenceRelationsRVsimpl}) that the series of missing terms is $0=a_1=a_4=a_7=\ldots=a_{3n+1}$ for both $R$ and $V$.

Note that our definition of the Airy functions, given above in Eqs.~(\ref{eq:OurAiryAi}) and~(\ref{eq:OurAiryBi}), is different from that in some mathematical literature; see, for instance,  Sec.~10.4 in~Ref.~\cite{BookAbramowitz1972}. In notations of Ref.~\cite{BookAbramowitz1972} our Airy functions $Ai(z)$ and $Bi(z)$ should be denoted as $f(z)$ and $g(z)$, respectively.

Finally, we present the complex plots of $R$ and $V$ based on selected values of their intrinsic parameters $\mu$, $\nu$ and $\lambda$, see Fig.~\ref{fig:RVAiBi}(b,c). We make the most obvious choice for the parameter values: $0$, $1$, and $i$. A notable observation here is the presence of a double zero at the origin for all the graphs of the $V$ function compared to the $R$ function and Airy functions.

\section{Conclusion} 
In summary, the Dirac equation for a confined massless fermion coupled to a gauge field, which is motivated by the eigenproblem of a half-bearded GNR under an external in-plane electric field, can be analytically solved using new $R$ and $V$ special functions. These functions are defined as power series, with coefficients determined recursively through Eq.~(\ref{eq:DiracEquation4GNRSingleComponentphiBEfieldRR1}). Similar to the Jackiw-Rebbi soliton in 2D~\cite{jackiw1976solitons}, the zero-energy modes of the given Dirac equation exhibit $1+1$ chiral anomaly favorable for scattering-free currents and highly sought for in the condensed matter physics within the context of topological phases of matter. As a striking contrast to other $1+1$ chiral anomalies~\cite{Arouca2022,frohlich_gauge_2023}, the chiral anomaly structure here exhibit a gauge field dependence, and therefore it is finely tunable. It is remarkable to observe that the configuration of the gauge field aligns with the potential current flow in the Hall effect: the field in our case is perpendicular to the transport direction. While further research is needed in this direction, this perpendicular configuration suggests reversibility between the anomaly and its inflow~\cite{callan_anomalies_1985,Witten2019}. We suggest additional investigation into the formal aspects of this anomaly as a quantum field anomaly.

\begin{acknowledgments}
The authors thank C.~Bennakker for the useful discussions in the initial equations of the half-bearded graphene nanoribbons. The authors acknowledge C.~A.~Downing and A.~Cipriani for their useful and critical comments. V.A.S. was partly supported by HORIZON EUROPE MSCA-2021-PF-01 (project no. 101065500, TeraExc). R.B.P. acknowledges the support from the Institute of Mathematical Sciences and Physics, University of the Philippines - Los Baños.

\end{acknowledgments}

\section*{Data Availability Statement}

The data that supports the findings of this study are available within the article.



\nocite{*}
\bibliography{library}

\begin{thebibliography}{45}%
\makeatletter
\providecommand \@ifxundefined [1]{%
 \@ifx{#1\undefined}
}%
\providecommand \@ifnum [1]{%
 \ifnum #1\expandafter \@firstoftwo
 \else \expandafter \@secondoftwo
 \fi
}%
\providecommand \@ifx [1]{%
 \ifx #1\expandafter \@firstoftwo
 \else \expandafter \@secondoftwo
 \fi
}%
\providecommand \natexlab [1]{#1}%
\providecommand \enquote  [1]{``#1''}%
\providecommand \bibnamefont  [1]{#1}%
\providecommand \bibfnamefont [1]{#1}%
\providecommand \citenamefont [1]{#1}%
\providecommand \href@noop [0]{\@secondoftwo}%
\providecommand \href [0]{\begingroup \@sanitize@url \@href}%
\providecommand \@href[1]{\@@startlink{#1}\@@href}%
\providecommand \@@href[1]{\endgroup#1\@@endlink}%
\providecommand \@sanitize@url [0]{\catcode `\\12\catcode `\$12\catcode `\&12\catcode `\#12\catcode `\^12\catcode `\_12\catcode `\%12\relax}%
\providecommand \@@startlink[1]{}%
\providecommand \@@endlink[0]{}%
\providecommand \url  [0]{\begingroup\@sanitize@url \@url }%
\providecommand \@url [1]{\endgroup\@href {#1}{\urlprefix }}%
\providecommand \urlprefix  [0]{URL }%
\providecommand \Eprint [0]{\href }%
\providecommand \doibase [0]{http://dx.doi.org/}%
\providecommand \selectlanguage [0]{\@gobble}%
\providecommand \bibinfo  [0]{\@secondoftwo}%
\providecommand \bibfield  [0]{\@secondoftwo}%
\providecommand \translation [1]{[#1]}%
\providecommand \BibitemOpen [0]{}%
\providecommand \bibitemStop [0]{}%
\providecommand \bibitemNoStop [0]{.\EOS\space}%
\providecommand \EOS [0]{\spacefactor3000\relax}%
\providecommand \BibitemShut  [1]{\csname bibitem#1\endcsname}%
\let\auto@bib@innerbib\@empty
\bibitem [{\citenamefont {Landau}\ and\ \citenamefont {Lifshitz}(2013)}]{landau2013quantum}%
  \BibitemOpen
  \bibfield  {author} {\bibinfo {author} {\bibfnamefont {L.~D.}\ \bibnamefont {Landau}}\ and\ \bibinfo {author} {\bibfnamefont {E.~M.}\ \bibnamefont {Lifshitz}},\ }\href@noop {} {\emph {\bibinfo {title} {Quantum mechanics: non-relativistic theory}}},\ Vol.~\bibinfo {volume} {3}\ (\bibinfo  {publisher} {Elsevier},\ \bibinfo {year} {2013})\BibitemShut {NoStop}%
\bibitem [{\citenamefont {Griffiths}\ and\ \citenamefont {Schroeter}(2018)}]{griffiths2018introduction}%
  \BibitemOpen
  \bibfield  {author} {\bibinfo {author} {\bibfnamefont {D.~J.}\ \bibnamefont {Griffiths}}\ and\ \bibinfo {author} {\bibfnamefont {D.~F.}\ \bibnamefont {Schroeter}},\ }\href@noop {} {\emph {\bibinfo {title} {Introduction to quantum mechanics}}}\ (\bibinfo  {publisher} {Cambridge university press},\ \bibinfo {year} {2018})\BibitemShut {NoStop}%
\bibitem [{\citenamefont {Yoshida}(1986)}]{yoshida1986classical}%
  \BibitemOpen
  \bibfield  {author} {\bibinfo {author} {\bibfnamefont {J.}~\bibnamefont {Yoshida}},\ }\bibfield  {title} {\enquote {\bibinfo {title} {Classical versus quantum mechanical calculation of the electron distribution at the n-algaas/gaas heterointerface},}\ }\href@noop {} {\bibfield  {journal} {\bibinfo  {journal} {IEEE transactions on electron devices}\ }\textbf {\bibinfo {volume} {33}},\ \bibinfo {pages} {154--156} (\bibinfo {year} {1986})}\BibitemShut {NoStop}%
\bibitem [{\citenamefont {Khondker}\ and\ \citenamefont {Anwar}(1987)}]{khondker1987analytical}%
  \BibitemOpen
  \bibfield  {author} {\bibinfo {author} {\bibfnamefont {A.}~\bibnamefont {Khondker}}\ and\ \bibinfo {author} {\bibfnamefont {A.}~\bibnamefont {Anwar}},\ }\bibfield  {title} {\enquote {\bibinfo {title} {Analytical models for algaas/gaas heterojunction quantum wells},}\ }\href@noop {} {\bibfield  {journal} {\bibinfo  {journal} {Solid-state electronics}\ }\textbf {\bibinfo {volume} {30}},\ \bibinfo {pages} {847--852} (\bibinfo {year} {1987})}\BibitemShut {NoStop}%
\bibitem [{\citenamefont {Zhang}\ and\ \citenamefont {Lynn}(1993)}]{zhang1993new}%
  \BibitemOpen
  \bibfield  {author} {\bibinfo {author} {\bibfnamefont {H.}~\bibnamefont {Zhang}}\ and\ \bibinfo {author} {\bibfnamefont {J.}~\bibnamefont {Lynn}},\ }\bibfield  {title} {\enquote {\bibinfo {title} {New exact solution of the one-dimensional schr{\"o}dinger equation and its application to polarized neutron reflectometry},}\ }\href@noop {} {\bibfield  {journal} {\bibinfo  {journal} {Physical review letters}\ }\textbf {\bibinfo {volume} {70}},\ \bibinfo {pages} {77} (\bibinfo {year} {1993})}\BibitemShut {NoStop}%
\bibitem [{\citenamefont {Parfitt}\ and\ \citenamefont {Portnoi}(2002)}]{parfitt2002two}%
  \BibitemOpen
  \bibfield  {author} {\bibinfo {author} {\bibfnamefont {D.}~\bibnamefont {Parfitt}}\ and\ \bibinfo {author} {\bibfnamefont {M.}~\bibnamefont {Portnoi}},\ }\bibfield  {title} {\enquote {\bibinfo {title} {The two-dimensional hydrogen atom revisited},}\ }\href@noop {} {\bibfield  {journal} {\bibinfo  {journal} {Journal of Mathematical Physics}\ }\textbf {\bibinfo {volume} {43}},\ \bibinfo {pages} {4681--4691} (\bibinfo {year} {2002})}\BibitemShut {NoStop}%
\bibitem [{\citenamefont {Lahiri}, \citenamefont {Roy},\ and\ \citenamefont {Bagchi}(1990)}]{lahiri1990supersymmetry}%
  \BibitemOpen
  \bibfield  {author} {\bibinfo {author} {\bibfnamefont {A.}~\bibnamefont {Lahiri}}, \bibinfo {author} {\bibfnamefont {P.~K.}\ \bibnamefont {Roy}}, \ and\ \bibinfo {author} {\bibfnamefont {B.}~\bibnamefont {Bagchi}},\ }\bibfield  {title} {\enquote {\bibinfo {title} {Supersymmetry in quantum mechanics},}\ }\href@noop {} {\bibfield  {journal} {\bibinfo  {journal} {International Journal of Modern Physics A}\ }\textbf {\bibinfo {volume} {5}},\ \bibinfo {pages} {1383--1456} (\bibinfo {year} {1990})}\BibitemShut {NoStop}%
\bibitem [{\citenamefont {Sokolov}(2008)}]{sokolov2008factorization}%
  \BibitemOpen
  \bibfield  {author} {\bibinfo {author} {\bibfnamefont {A.}~\bibnamefont {Sokolov}},\ }\bibfield  {title} {\enquote {\bibinfo {title} {Factorization of nonlinear supersymmetry in one-dimensional quantum mechanics. ii: Proofs of theorems on reducibility},}\ }\href@noop {} {\bibfield  {journal} {\bibinfo  {journal} {Journal of Mathematical Sciences}\ }\textbf {\bibinfo {volume} {151}},\ \bibinfo {pages} {2924--2936} (\bibinfo {year} {2008})}\BibitemShut {NoStop}%
\bibitem [{\citenamefont {Downing}(2013)}]{downing2013solution}%
  \BibitemOpen
  \bibfield  {author} {\bibinfo {author} {\bibfnamefont {C.}~\bibnamefont {Downing}},\ }\bibfield  {title} {\enquote {\bibinfo {title} {On a solution of the schr{\"o}dinger equation with a hyperbolic double-well potential},}\ }\href@noop {} {\bibfield  {journal} {\bibinfo  {journal} {Journal of Mathematical Physics}\ }\textbf {\bibinfo {volume} {54}} (\bibinfo {year} {2013})}\BibitemShut {NoStop}%
\bibitem [{\citenamefont {Hartmann}(2014)}]{hartmann2014bound}%
  \BibitemOpen
  \bibfield  {author} {\bibinfo {author} {\bibfnamefont {R.~R.}\ \bibnamefont {Hartmann}},\ }\bibfield  {title} {\enquote {\bibinfo {title} {Bound states in a hyperbolic asymmetric double-well},}\ }\href@noop {} {\bibfield  {journal} {\bibinfo  {journal} {Journal of Mathematical Physics}\ }\textbf {\bibinfo {volume} {55}} (\bibinfo {year} {2014})}\BibitemShut {NoStop}%
\bibitem [{\citenamefont {Manning}(1935)}]{manning1935exact}%
  \BibitemOpen
  \bibfield  {author} {\bibinfo {author} {\bibfnamefont {M.~F.}\ \bibnamefont {Manning}},\ }\bibfield  {title} {\enquote {\bibinfo {title} {Exact solutions of the schr{\"o}dinger equation},}\ }\href@noop {} {\bibfield  {journal} {\bibinfo  {journal} {Physical Review}\ }\textbf {\bibinfo {volume} {48}},\ \bibinfo {pages} {161} (\bibinfo {year} {1935})}\BibitemShut {NoStop}%
\bibitem [{\citenamefont {Shahnazaryan}\ \emph {et~al.}(2014)\citenamefont {Shahnazaryan}, \citenamefont {Ishkhanyan}, \citenamefont {Shahverdyan},\ and\ \citenamefont {Ishkhanyan}}]{shahnazaryan2014new}%
  \BibitemOpen
  \bibfield  {author} {\bibinfo {author} {\bibfnamefont {V.}~\bibnamefont {Shahnazaryan}}, \bibinfo {author} {\bibfnamefont {T.}~\bibnamefont {Ishkhanyan}}, \bibinfo {author} {\bibfnamefont {T.}~\bibnamefont {Shahverdyan}}, \ and\ \bibinfo {author} {\bibfnamefont {A.}~\bibnamefont {Ishkhanyan}},\ }\bibfield  {title} {\enquote {\bibinfo {title} {New relations for the derivative of the confluent heun function},}\ }\href@noop {} {\bibfield  {journal} {\bibinfo  {journal} {arXiv preprint arXiv:1402.1318}\ } (\bibinfo {year} {2014})}\BibitemShut {NoStop}%
\bibitem [{\citenamefont {Ishkhanyan}\ and\ \citenamefont {Ishkhanyan}(2017)}]{ishkhanyan2017}%
  \BibitemOpen
  \bibfield  {author} {\bibinfo {author} {\bibfnamefont {T.}~\bibnamefont {Ishkhanyan}}\ and\ \bibinfo {author} {\bibfnamefont {A.}~\bibnamefont {Ishkhanyan}},\ }\bibfield  {title} {\enquote {\bibinfo {title} {Solutions of the bi-confluent heun equation in terms of the hermite functions},}\ }\href {\doibase https://doi.org/10.1016/j.aop.2017.04.015} {\bibfield  {journal} {\bibinfo  {journal} {Annals of Physics}\ }\textbf {\bibinfo {volume} {383}},\ \bibinfo {pages} {79--91} (\bibinfo {year} {2017})}\BibitemShut {NoStop}%
\bibitem [{\citenamefont {Abe}\ and\ \citenamefont {Fujita}(1987)}]{abe1987simple}%
  \BibitemOpen
  \bibfield  {author} {\bibinfo {author} {\bibfnamefont {S.}~\bibnamefont {Abe}}\ and\ \bibinfo {author} {\bibfnamefont {T.}~\bibnamefont {Fujita}},\ }\bibfield  {title} {\enquote {\bibinfo {title} {A simple analytic solution of the dirac equation with a scalar linear potential},}\ }\href@noop {} {\bibfield  {journal} {\bibinfo  {journal} {Nuclear Physics A}\ }\textbf {\bibinfo {volume} {475}},\ \bibinfo {pages} {657--662} (\bibinfo {year} {1987})}\BibitemShut {NoStop}%
\bibitem [{\citenamefont {Hofer}\ and\ \citenamefont {Stocker}(1989)}]{hofer1989dirac}%
  \BibitemOpen
  \bibfield  {author} {\bibinfo {author} {\bibfnamefont {D.}~\bibnamefont {Hofer}}\ and\ \bibinfo {author} {\bibfnamefont {W.}~\bibnamefont {Stocker}},\ }\bibfield  {title} {\enquote {\bibinfo {title} {A dirac particle in a scalar potential depending on one spatial coordinate},}\ }\href@noop {} {\bibfield  {journal} {\bibinfo  {journal} {Physics Letters A}\ }\textbf {\bibinfo {volume} {138}},\ \bibinfo {pages} {463--464} (\bibinfo {year} {1989})}\BibitemShut {NoStop}%
\bibitem [{\citenamefont {Chargui}, \citenamefont {Trabelsi},\ and\ \citenamefont {Chetouani}(2010)}]{chargui2010exact}%
  \BibitemOpen
  \bibfield  {author} {\bibinfo {author} {\bibfnamefont {Y.}~\bibnamefont {Chargui}}, \bibinfo {author} {\bibfnamefont {A.}~\bibnamefont {Trabelsi}}, \ and\ \bibinfo {author} {\bibfnamefont {L.}~\bibnamefont {Chetouani}},\ }\bibfield  {title} {\enquote {\bibinfo {title} {Exact solution of the (1+ 1)-dimensional dirac equation with vector and scalar linear potentials in the presence of a minimal length},}\ }\href@noop {} {\bibfield  {journal} {\bibinfo  {journal} {Physics Letters A}\ }\textbf {\bibinfo {volume} {374}},\ \bibinfo {pages} {531--534} (\bibinfo {year} {2010})}\BibitemShut {NoStop}%
\bibitem [{\citenamefont {Tezuka}(2013)}]{tezuka2013analytical}%
  \BibitemOpen
  \bibfield  {author} {\bibinfo {author} {\bibfnamefont {H.}~\bibnamefont {Tezuka}},\ }\bibfield  {title} {\enquote {\bibinfo {title} {Analytical solutions of the dirac equation with a scalar linear potential},}\ }\href@noop {} {\bibfield  {journal} {\bibinfo  {journal} {AIP Advances}\ }\textbf {\bibinfo {volume} {3}} (\bibinfo {year} {2013})}\BibitemShut {NoStop}%
\bibitem [{\citenamefont {Jaronski}(2021)}]{jaronski2021linear}%
  \BibitemOpen
  \bibfield  {author} {\bibinfo {author} {\bibfnamefont {W.~S.}\ \bibnamefont {Jaronski}},\ }\bibfield  {title} {\enquote {\bibinfo {title} {The linear potential and the dirac equation},}\ }\href@noop {} {\bibfield  {journal} {\bibinfo  {journal} {arXiv preprint arXiv:2108.05953}\ } (\bibinfo {year} {2021})}\BibitemShut {NoStop}%
\bibitem [{\citenamefont {Jackiw}\ and\ \citenamefont {Semenoff}(1983)}]{jackiw1983continuum}%
  \BibitemOpen
  \bibfield  {author} {\bibinfo {author} {\bibfnamefont {R.}~\bibnamefont {Jackiw}}\ and\ \bibinfo {author} {\bibfnamefont {G.}~\bibnamefont {Semenoff}},\ }\bibfield  {title} {\enquote {\bibinfo {title} {Continuum quantum field theory for a linearly conjugated diatomic polymer with fermion fractionization},}\ }\href@noop {} {\bibfield  {journal} {\bibinfo  {journal} {Physical Review Letters}\ }\textbf {\bibinfo {volume} {50}},\ \bibinfo {pages} {439} (\bibinfo {year} {1983})}\BibitemShut {NoStop}%
\bibitem [{\citenamefont {DiVincenzo}\ and\ \citenamefont {Mele}(1984)}]{divincenzo1984self}%
  \BibitemOpen
  \bibfield  {author} {\bibinfo {author} {\bibfnamefont {D.}~\bibnamefont {DiVincenzo}}\ and\ \bibinfo {author} {\bibfnamefont {E.}~\bibnamefont {Mele}},\ }\bibfield  {title} {\enquote {\bibinfo {title} {Self-consistent effective-mass theory for intralayer screening in graphite intercalation compounds},}\ }\href@noop {} {\bibfield  {journal} {\bibinfo  {journal} {Physical Review B}\ }\textbf {\bibinfo {volume} {29}},\ \bibinfo {pages} {1685} (\bibinfo {year} {1984})}\BibitemShut {NoStop}%
\bibitem [{\citenamefont {Renan}, \citenamefont {Pacheco},\ and\ \citenamefont {Almeida}(2000)}]{renan2000treating}%
  \BibitemOpen
  \bibfield  {author} {\bibinfo {author} {\bibfnamefont {R.}~\bibnamefont {Renan}}, \bibinfo {author} {\bibfnamefont {M.}~\bibnamefont {Pacheco}}, \ and\ \bibinfo {author} {\bibfnamefont {C.}~\bibnamefont {Almeida}},\ }\bibfield  {title} {\enquote {\bibinfo {title} {Treating some solid state problems with the dirac equation},}\ }\href@noop {} {\bibfield  {journal} {\bibinfo  {journal} {Journal of Physics A: Mathematical and General}\ }\textbf {\bibinfo {volume} {33}},\ \bibinfo {pages} {L509} (\bibinfo {year} {2000})}\BibitemShut {NoStop}%
\bibitem [{\citenamefont {Geim}\ and\ \citenamefont {Novoselov}(2007)}]{geim2007rise}%
  \BibitemOpen
  \bibfield  {author} {\bibinfo {author} {\bibfnamefont {A.~K.}\ \bibnamefont {Geim}}\ and\ \bibinfo {author} {\bibfnamefont {K.~S.}\ \bibnamefont {Novoselov}},\ }\bibfield  {title} {\enquote {\bibinfo {title} {The rise of graphene},}\ }\href@noop {} {\bibfield  {journal} {\bibinfo  {journal} {Nature materials}\ }\textbf {\bibinfo {volume} {6}},\ \bibinfo {pages} {183--191} (\bibinfo {year} {2007})}\BibitemShut {NoStop}%
\bibitem [{\citenamefont {Bergman}\ and\ \citenamefont {Le~Hur}(2009)}]{bergman2009near}%
  \BibitemOpen
  \bibfield  {author} {\bibinfo {author} {\bibfnamefont {D.~L.}\ \bibnamefont {Bergman}}\ and\ \bibinfo {author} {\bibfnamefont {K.}~\bibnamefont {Le~Hur}},\ }\bibfield  {title} {\enquote {\bibinfo {title} {Near-zero modes in condensate phases of the dirac theory on the honeycomb lattice},}\ }\href@noop {} {\bibfield  {journal} {\bibinfo  {journal} {Physical Review B}\ }\textbf {\bibinfo {volume} {79}},\ \bibinfo {pages} {184520} (\bibinfo {year} {2009})}\BibitemShut {NoStop}%
\bibitem [{\citenamefont {Brey}\ and\ \citenamefont {Fertig}(2006)}]{brey2006electronic}%
  \BibitemOpen
  \bibfield  {author} {\bibinfo {author} {\bibfnamefont {L.}~\bibnamefont {Brey}}\ and\ \bibinfo {author} {\bibfnamefont {H.}~\bibnamefont {Fertig}},\ }\bibfield  {title} {\enquote {\bibinfo {title} {Electronic states of graphene nanoribbons studied with the dirac equation},}\ }\href@noop {} {\bibfield  {journal} {\bibinfo  {journal} {Physical Review B}\ }\textbf {\bibinfo {volume} {73}},\ \bibinfo {pages} {235411} (\bibinfo {year} {2006})}\BibitemShut {NoStop}%
\bibitem [{\citenamefont {Akhmerov}\ and\ \citenamefont {Beenakker}(2008)}]{akhmerov2008boundary}%
  \BibitemOpen
  \bibfield  {author} {\bibinfo {author} {\bibfnamefont {A.}~\bibnamefont {Akhmerov}}\ and\ \bibinfo {author} {\bibfnamefont {C.}~\bibnamefont {Beenakker}},\ }\bibfield  {title} {\enquote {\bibinfo {title} {Boundary conditions for dirac fermions on a terminated honeycomb lattice},}\ }\href@noop {} {\bibfield  {journal} {\bibinfo  {journal} {Physical Review B}\ }\textbf {\bibinfo {volume} {77}},\ \bibinfo {pages} {085423} (\bibinfo {year} {2008})}\BibitemShut {NoStop}%
\bibitem [{\citenamefont {Wakabayashi}\ \emph {et~al.}(2010)\citenamefont {Wakabayashi}, \citenamefont {Okada}, \citenamefont {Tomita}, \citenamefont {Fujimoto},\ and\ \citenamefont {Natsume}}]{wakabayashi2010edge}%
  \BibitemOpen
  \bibfield  {author} {\bibinfo {author} {\bibfnamefont {K.}~\bibnamefont {Wakabayashi}}, \bibinfo {author} {\bibfnamefont {S.}~\bibnamefont {Okada}}, \bibinfo {author} {\bibfnamefont {R.}~\bibnamefont {Tomita}}, \bibinfo {author} {\bibfnamefont {S.}~\bibnamefont {Fujimoto}}, \ and\ \bibinfo {author} {\bibfnamefont {Y.}~\bibnamefont {Natsume}},\ }\bibfield  {title} {\enquote {\bibinfo {title} {Edge states and flat bands of graphene nanoribbons with edge modification},}\ }\href@noop {} {\bibfield  {journal} {\bibinfo  {journal} {Journal of the Physical Society of Japan}\ }\textbf {\bibinfo {volume} {79}},\ \bibinfo {pages} {034706} (\bibinfo {year} {2010})}\BibitemShut {NoStop}%
\bibitem [{\citenamefont {Jaskólski}\ \emph {et~al.}(2011)\citenamefont {Jaskólski}, \citenamefont {Ayuela}, \citenamefont {Pelc}, \citenamefont {Santos},\ and\ \citenamefont {Chico}}]{Jaskolski2011}%
  \BibitemOpen
  \bibfield  {author} {\bibinfo {author} {\bibfnamefont {W.}~\bibnamefont {Jaskólski}}, \bibinfo {author} {\bibfnamefont {A.}~\bibnamefont {Ayuela}}, \bibinfo {author} {\bibfnamefont {M.}~\bibnamefont {Pelc}}, \bibinfo {author} {\bibfnamefont {H.}~\bibnamefont {Santos}}, \ and\ \bibinfo {author} {\bibfnamefont {L.}~\bibnamefont {Chico}},\ }\bibfield  {title} {\enquote {\bibinfo {title} {Edge states and flat bands in graphene nanoribbons with arbitrary geometries},}\ }\href {\doibase 10.1103/PhysRevB.83.235424} {\bibfield  {journal} {\bibinfo  {journal} {Physical Review B}\ }\textbf {\bibinfo {volume} {83}},\ \bibinfo {pages} {235424} (\bibinfo {year} {2011})}\BibitemShut {NoStop}%
\bibitem [{\citenamefont {Li}\ \emph {et~al.}(2012)\citenamefont {Li}, \citenamefont {Lin}, \citenamefont {Chang},\ and\ \citenamefont {Lin}}]{li2012electronic}%
  \BibitemOpen
  \bibfield  {author} {\bibinfo {author} {\bibfnamefont {T.}~\bibnamefont {Li}}, \bibinfo {author} {\bibfnamefont {M.}~\bibnamefont {Lin}}, \bibinfo {author} {\bibfnamefont {S.}~\bibnamefont {Chang}}, \ and\ \bibinfo {author} {\bibfnamefont {T.}~\bibnamefont {Lin}},\ }\bibfield  {title} {\enquote {\bibinfo {title} {Electronic properties of bearded graphene nanoribbons},}\ }\href@noop {} {\bibfield  {journal} {\bibinfo  {journal} {Journal of Physics and Chemistry of Solids}\ }\textbf {\bibinfo {volume} {73}},\ \bibinfo {pages} {1245--1251} (\bibinfo {year} {2012})}\BibitemShut {NoStop}%
\bibitem [{\citenamefont {Saroka}\ \emph {et~al.}(2015)\citenamefont {Saroka}, \citenamefont {Batrakov}, \citenamefont {Demin},\ and\ \citenamefont {Chernozatonskii}}]{Saroka2015}%
  \BibitemOpen
  \bibfield  {author} {\bibinfo {author} {\bibfnamefont {V.~A.}\ \bibnamefont {Saroka}}, \bibinfo {author} {\bibfnamefont {K.~G.}\ \bibnamefont {Batrakov}}, \bibinfo {author} {\bibfnamefont {V.~A.}\ \bibnamefont {Demin}}, \ and\ \bibinfo {author} {\bibfnamefont {L.~A.}\ \bibnamefont {Chernozatonskii}},\ }\bibfield  {title} {\enquote {\bibinfo {title} {Band gaps in jagged and straight graphene nanoribbons tunable by an external electric field},}\ }\href {\doibase 10.1088/0953-8984/27/14/145305} {\bibfield  {journal} {\bibinfo  {journal} {Journal of physics. Condensed matter : an Institute of Physics journal}\ }\textbf {\bibinfo {volume} {27}},\ \bibinfo {pages} {145305} (\bibinfo {year} {2015})}\BibitemShut {NoStop}%
\bibitem [{\citenamefont {Saroka}\ and\ \citenamefont {Batrakov}(2015)}]{Saroka2015a}%
  \BibitemOpen
  \bibfield  {author} {\bibinfo {author} {\bibfnamefont {V.~A.}\ \bibnamefont {Saroka}}\ and\ \bibinfo {author} {\bibfnamefont {K.~G.}\ \bibnamefont {Batrakov}},\ }\bibfield  {title} {\enquote {\bibinfo {title} {Dirac electrons of graphene nanoribbons tunable by transverse electric field},}\ }in\ \href {\doibase 10.1142/9789814696524_0060} {\emph {\bibinfo {booktitle} {Physics, {Chemistry} and {Applications} of {Nanostructures}}}},\ \bibinfo {editor} {edited by\ \bibinfo {editor} {\bibfnamefont {V.~E.}\ \bibnamefont {Borisenko}}, \bibinfo {editor} {\bibfnamefont {S.~V.}\ \bibnamefont {Gaponenko}}, \bibinfo {editor} {\bibfnamefont {V.~S.}\ \bibnamefont {Gurin}}, \ and\ \bibinfo {editor} {\bibfnamefont {C.~H.}\ \bibnamefont {Kam}}}\ (\bibinfo  {publisher} {World Scientific},\ \bibinfo {year} {2015})\ pp.\ \bibinfo {pages} {240--243}\BibitemShut {NoStop}%
\bibitem [{\citenamefont {Saroka}\ and\ \citenamefont {Batrakov}(2016)}]{Saroka2016a}%
  \BibitemOpen
  \bibfield  {author} {\bibinfo {author} {\bibfnamefont {V.~A.}\ \bibnamefont {Saroka}}\ and\ \bibinfo {author} {\bibfnamefont {K.~G.}\ \bibnamefont {Batrakov}},\ }\bibfield  {title} {\enquote {\bibinfo {title} {Zigzag-shaped superlattices on the basis of graphene nanoribbons: {Structure} and electronic properties},}\ }\href {\doibase 10.1007/s11182-016-0816-6} {\bibfield  {journal} {\bibinfo  {journal} {Russian Physics Journal}\ }\textbf {\bibinfo {volume} {59}},\ \bibinfo {pages} {633--639} (\bibinfo {year} {2016})}\BibitemShut {NoStop}%
\bibitem [{\citenamefont {Wakabayashi}(2001)}]{Wakabayashi2001}%
  \BibitemOpen
  \bibfield  {author} {\bibinfo {author} {\bibfnamefont {K.}~\bibnamefont {Wakabayashi}},\ }\bibfield  {title} {\enquote {\bibinfo {title} {Electronic transport properties of nanographite ribbon junctions},}\ }\href {\doibase 10.1103/PhysRevB.64.125428} {\bibfield  {journal} {\bibinfo  {journal} {Physical Review B}\ }\textbf {\bibinfo {volume} {64}},\ \bibinfo {pages} {125428} (\bibinfo {year} {2001})}\BibitemShut {NoStop}%
\bibitem [{\citenamefont {Kusakabe}\ and\ \citenamefont {Maruyama}(2003)}]{Kusakabe2003}%
  \BibitemOpen
  \bibfield  {author} {\bibinfo {author} {\bibfnamefont {K.}~\bibnamefont {Kusakabe}}\ and\ \bibinfo {author} {\bibfnamefont {M.}~\bibnamefont {Maruyama}},\ }\bibfield  {title} {\enquote {\bibinfo {title} {Magnetic nanographite},}\ }\href {\doibase 10.1103/PhysRevB.67.092406} {\bibfield  {journal} {\bibinfo  {journal} {Physical Review B}\ }\textbf {\bibinfo {volume} {67}},\ \bibinfo {pages} {092406} (\bibinfo {year} {2003})}\BibitemShut {NoStop}%
\bibitem [{\citenamefont {Kohmoto}\ and\ \citenamefont {Hasegawa}(2007)}]{Kohmoto2007}%
  \BibitemOpen
  \bibfield  {author} {\bibinfo {author} {\bibfnamefont {M.}~\bibnamefont {Kohmoto}}\ and\ \bibinfo {author} {\bibfnamefont {Y.}~\bibnamefont {Hasegawa}},\ }\bibfield  {title} {\enquote {\bibinfo {title} {Zero modes and edge states of the honeycomb lattice},}\ }\href {\doibase 10.1103/PhysRevB.76.205402} {\bibfield  {journal} {\bibinfo  {journal} {Physical Review B}\ }\textbf {\bibinfo {volume} {76}},\ \bibinfo {pages} {205402} (\bibinfo {year} {2007})}\BibitemShut {NoStop}%
\bibitem [{\citenamefont {Saroka}\ \emph {et~al.}(2023)\citenamefont {Saroka}, \citenamefont {Kong}, \citenamefont {Downing}, \citenamefont {Payod}, \citenamefont {Fischer}, \citenamefont {Sun},\ and\ \citenamefont {Bogani}}]{saroka2023tunable}%
  \BibitemOpen
  \bibfield  {author} {\bibinfo {author} {\bibfnamefont {V.~A.}\ \bibnamefont {Saroka}}, \bibinfo {author} {\bibfnamefont {F.}~\bibnamefont {Kong}}, \bibinfo {author} {\bibfnamefont {C.~A.}\ \bibnamefont {Downing}}, \bibinfo {author} {\bibfnamefont {R.~B.}\ \bibnamefont {Payod}}, \bibinfo {author} {\bibfnamefont {F.~R.}\ \bibnamefont {Fischer}}, \bibinfo {author} {\bibfnamefont {X.}~\bibnamefont {Sun}}, \ and\ \bibinfo {author} {\bibfnamefont {L.}~\bibnamefont {Bogani}},\ }\bibfield  {title} {\enquote {\bibinfo {title} {Tunable chiral anomalies and coherent transport on a honeycomb lattice},}\ }\href@noop {} {\bibfield  {journal} {\bibinfo  {journal} {arXiv preprint arXiv:2310.02148}\ } (\bibinfo {year} {2023})}\BibitemShut {NoStop}%
\bibitem [{\citenamefont {Landau}\ and\ \citenamefont {Lifshitz}(1975)}]{landau1975classical}%
  \BibitemOpen
  \bibfield  {author} {\bibinfo {author} {\bibfnamefont {L.}~\bibnamefont {Landau}}\ and\ \bibinfo {author} {\bibfnamefont {E.}~\bibnamefont {Lifshitz}},\ }\href {https://books.google.com.ph/books?id=X18PF4oKyrUC} {\emph {\bibinfo {title} {The Classical Theory of Fields: Volume 2}}},\ Course of theoretical physics\ (\bibinfo  {publisher} {Elsevier Science},\ \bibinfo {year} {1975})\BibitemShut {NoStop}%
\bibitem [{\citenamefont {Berestetskii}, \citenamefont {Lifshitz},\ and\ \citenamefont {Pitaevskii}(1982)}]{berestetskii1982quantum}%
  \BibitemOpen
  \bibfield  {author} {\bibinfo {author} {\bibfnamefont {V.}~\bibnamefont {Berestetskii}}, \bibinfo {author} {\bibfnamefont {E.}~\bibnamefont {Lifshitz}}, \ and\ \bibinfo {author} {\bibfnamefont {L.}~\bibnamefont {Pitaevskii}},\ }\href {https://books.google.com.ph/books?id=YlwKR5JNWDgC} {\emph {\bibinfo {title} {Quantum Electrodynamics: Volume 4}}},\ Course of theoretical physics\ (\bibinfo  {publisher} {Elsevier Science},\ \bibinfo {year} {1982})\BibitemShut {NoStop}%
\bibitem [{\citenamefont {Kreyszig}(2020)}]{kreyszig2020advanced}%
  \BibitemOpen
  \bibfield  {author} {\bibinfo {author} {\bibfnamefont {E.}~\bibnamefont {Kreyszig}},\ }\href {https://books.google.com.ph/books?id=w4T3DwAAQBAJ} {\emph {\bibinfo {title} {Advanced Engineering Mathematics}}},\ \bibinfo {edition} {10th}\ ed.\ (\bibinfo  {publisher} {Wiley},\ \bibinfo {year} {2020})\BibitemShut {NoStop}%
\bibitem [{\citenamefont {Jackiw}\ and\ \citenamefont {Rebbi}(1976)}]{jackiw1976solitons}%
  \BibitemOpen
  \bibfield  {author} {\bibinfo {author} {\bibfnamefont {R.}~\bibnamefont {Jackiw}}\ and\ \bibinfo {author} {\bibfnamefont {C.}~\bibnamefont {Rebbi}},\ }\bibfield  {title} {\enquote {\bibinfo {title} {Solitons with fermion number $1/2$},}\ }\href@noop {} {\bibfield  {journal} {\bibinfo  {journal} {Physical Review D}\ }\textbf {\bibinfo {volume} {13}},\ \bibinfo {pages} {3398} (\bibinfo {year} {1976})}\BibitemShut {NoStop}%
\bibitem [{\citenamefont {Shen}(2012)}]{shen2012topological}%
  \BibitemOpen
  \bibfield  {author} {\bibinfo {author} {\bibfnamefont {S.-Q.}\ \bibnamefont {Shen}},\ }\href@noop {} {\emph {\bibinfo {title} {Topological insulators}}},\ Vol.\ \bibinfo {volume} {174}\ (\bibinfo  {publisher} {Springer},\ \bibinfo {year} {2012})\BibitemShut {NoStop}%
\bibitem [{\citenamefont {Abramowitz}\ and\ \citenamefont {Stegun}(1972)}]{BookAbramowitz1972}%
  \BibitemOpen
  \bibfield  {author} {\bibinfo {author} {\bibfnamefont {M.}~\bibnamefont {Abramowitz}}\ and\ \bibinfo {author} {\bibfnamefont {I.~A.}\ \bibnamefont {Stegun}},\ }\href@noop {} {\emph {\bibinfo {title} {Handbook of Mathematical Functions With Formulas, Graphs and Mathematical Tables}}},\ \bibinfo {edition} {10th}\ ed.,\ \bibinfo {series} {Applied mathematics series}\ No.~\bibinfo {number} {55}\ (\bibinfo  {publisher} {United States. Government Printing Office},\ \bibinfo {year} {1972})\BibitemShut {NoStop}%
\bibitem [{\citenamefont {Arouca}, \citenamefont {Cappelli},\ and\ \citenamefont {Hansson}(2022)}]{Arouca2022}%
  \BibitemOpen
  \bibfield  {author} {\bibinfo {author} {\bibfnamefont {R.}~\bibnamefont {Arouca}}, \bibinfo {author} {\bibfnamefont {A.}~\bibnamefont {Cappelli}}, \ and\ \bibinfo {author} {\bibfnamefont {T.~H.}\ \bibnamefont {Hansson}},\ }\bibfield  {title} {\enquote {\bibinfo {title} {Quantum field theory anomalies in condensed matter physics},}\ }\href {\doibase 10.21468/SciPostPhysLectNotes.62} {\bibfield  {journal} {\bibinfo  {journal} {SciPost Phys. Lect. Notes}\ ,\ \bibinfo {pages} {62}} (\bibinfo {year} {2022})}\BibitemShut {NoStop}%
\bibitem [{\citenamefont {Fröhlich}(2023)}]{frohlich_gauge_2023}%
  \BibitemOpen
  \bibfield  {author} {\bibinfo {author} {\bibfnamefont {J.}~\bibnamefont {Fröhlich}},\ }\bibfield  {title} {\enquote {\bibinfo {title} {Gauge invariance and anomalies in condensed matter physics},}\ }\href {\doibase 10.1063/5.0135142} {\bibfield  {journal} {\bibinfo  {journal} {Journal of Mathematical Physics}\ }\textbf {\bibinfo {volume} {64}},\ \bibinfo {pages} {031903} (\bibinfo {year} {2023})}\BibitemShut {NoStop}%
\bibitem [{\citenamefont {Callan}\ and\ \citenamefont {Harvey}(1985)}]{callan_anomalies_1985}%
  \BibitemOpen
  \bibfield  {author} {\bibinfo {author} {\bibfnamefont {C.}~\bibnamefont {Callan}}\ and\ \bibinfo {author} {\bibfnamefont {J.}~\bibnamefont {Harvey}},\ }\bibfield  {title} {\enquote {\bibinfo {title} {Anomalies and fermion zero modes on strings and domain walls},}\ }\href {\doibase 10.1016/0550-3213(85)90489-4} {\bibfield  {journal} {\bibinfo  {journal} {Nuclear Physics B}\ }\textbf {\bibinfo {volume} {250}},\ \bibinfo {pages} {427--436} (\bibinfo {year} {1985})}\BibitemShut {NoStop}%
\bibitem [{\citenamefont {Witten}\ and\ \citenamefont {Yonekura}(2019)}]{Witten2019}%
  \BibitemOpen
  \bibfield  {author} {\bibinfo {author} {\bibfnamefont {E.}~\bibnamefont {Witten}}\ and\ \bibinfo {author} {\bibfnamefont {K.}~\bibnamefont {Yonekura}},\ }\href {http://arxiv.org/abs/1909.08775} {\enquote {\bibinfo {title} {Anomaly {Inflow} and the $\eta$-{Invariant}},}\ } (\bibinfo {year} {2019}),\ \bibinfo {note} {arXiv: 1909.08775}\BibitemShut {NoStop}%
\end{thebibliography}%

\end{document}